\documentclass[aps,prb,twocolumn,showpacs,groupedaddress]{revtex4}

\usepackage{graphicx}

\begin{document}

\preprint{}

\title{Metallic state in La-doped YBa$_2$Cu$_3$O$_y$ thin films with $n$-type charge carriers}

\author{S. W. Zeng$^{1,2}$}

\author{X. Wang$^{1,2}$}

\author{W. M. L\"{u}$^{1,3}$}

\author{Z. Huang$^{1}$}

\author{M. Motapothula$^{1,2}$}

\author{Z. Q. Liu$^{1,2}$}

\author{Y. L. Zhao$^{1,2}$}

\author{A. Annadi$^{1,2}$}

\author{S. Dhar$^{1,3}$}

\author{H. Mao$^{4}$}

\author{W. Chen$^{2,4}$}

\author{T. Venkatesan$^{1,2,3}$}

\author{Ariando$^{1,2}$}

\altaffiliation[Email: ]{ariando@nus.edu.sg}

\affiliation{$^1$NUSNNI-Nanocore, $^2$Department of Physics, $^3$Department of Electrical and Computer Engineering, $^4$Department of Chemistry, National University of Singapore, Singapore}

\begin{abstract}
We report hole and electron doping in La-doped YBa$_2$Cu$_3$O$_y$
(YBCO) thin films synthesized by pulsed laser deposition technique
and subsequent \emph{in-situ} postannealing in oxygen ambient and
vaccum. The $n$-type samples show a metallic behavior below the Mott
limit and a high carrier density of $\sim2.8$ $\times$ 10$^{21}$
cm$^{-3}$ at room temperature (\emph{T}) at the optimally reduced
condition. The in-plane resistivity ( $\rho$$_{ab}$ ) of the
$n$-type samples exhibits a quadratic \emph{T} dependence in the
moderate-\emph{T} range and shows an anomaly at a relatively higher
\emph{T} probably related to pseudogap formation analogous to
underdoped Nd$_{2-x}$Ce$_x$CuO$_4$ (NCCO). Furthermore,
$\rho$$_{ab}$(T), \emph{T}$_c$ and \emph{T} with minimum resistivity
(\emph{T}$_{min}$) were investigated in both $p$- and $n$-side. The
present results reveal the $n$-$p$ asymmetry (symmetry) within the
metallic-state region in an underdoped cuprate and suggest the
potential toward ambipolar superconductivity in a single YBCO
system.
\end{abstract}



\maketitle
\section{Introduction}

High-\emph{T}$_c$ superconductivity in cuprates results from doping
charge carriers into Mott insulators. The magnetic and
superconducting properties of Mott insulators doped by holes
($p$-type) are different from those doped by electrons ($n$-type),
showing asymmetry in the phase diagrams [1]. Moreover, their
transport properties depend on the type of carriers, for example,
in-plane normal-state resistivity exhibits quadratic temperature
(\emph{T}) dependence for $n$-type cuprates [2] while linear
\emph{T} dependence is seen for $p$-type cuprates with optimal
doping [3,4]. Such comparisons of electron- and hole-doping
asymmetry (symmetry) in cuprates should help to further our
understanding of the cuprate superconductors [5,6]. The typical
electron-hole asymmetry (symmetry) investigation is usually based on
the cuprates with different crystallographic structure such as NCCO
(T' structure) and La$_{2-x}$Sr$_x$CuO$_4$ (T structure). These
materials have different parent Mott insulators, and thus, exhibit
different properties even without doping [7]. Therefore, it is
desirable to dope electrons and holes into a Mott insulator without
changing the crystallographic structure and address the physics in
both sides of such `ambipolar' cuprate.

An ambipolar single-crystal cuprate was first achieved in
Y$_{1-z}$La$_z$(Ba$_{1-x}$La$_x$)$_2$Cu$_3$O$_y$ (x=0.13, z=0.62)
with charge carriers ranging from $7\%$ of holes per Cu to $2\%$ of
electrons per Cu [8]. It was also found that
Y$_{0.38}$La$_{0.62}$(Ba$_{0.87}$La$_{0.13}$)$_2$Cu$_3$O$_y$
exhibits asymmetric behavior in magnetic ground states and transport
properties between electron- and hole-doped sides [9]. However, this
asymmetric comparison was limited to insulating state near the
zero-doping region due to low carrier density in $n$-type samples.
In order to fully investigate the $n$-$p$ asymmetry (symmetry), more
heavily electron-doped materials in YBCO system, which exhibit
metallic and even superconducting behaviors, are desirable.

Effective reduction of the as-grown materials is necessary to dope
$n$-type carriers in cuprates [10]. In general, the bulk single
crystals are annealed at high \emph{T} (850 $^\circ$C-1080
$^\circ$C) in low oxygen partial pressure (\emph{P}$_{O2}$) for tens
of hours to several days. For the thin-film growth, oxygen can be
removed uniformly and efficiently with postannealing for several
tens of minutes, since oxygen diffusion along the c-axis is much
easier and the diffusion lengths are comparable to the film
thickness [1]. Recently, an electrochemical technique was used to
remove oxygen in pure YBCO thin films and achieve $n$-type metallic
state [11]. However, carrier density of the $n$-type sample was
still low ($\sim2.5$ $\times$ 10$^{20}$ cm$^{-3}$) and the metallic
state was only observed above 120 K. Chemical doping is necessary
for achieving higher electron density. Since La$^{3+}$ possesses a
higher valence state than Ba$^{2+}$, La substitution for Ba in YBCO
is expected to provide additional electrons [8]. Therefore, one can
expect that more electrons can be doped by varying the La content in
Y$_{1-z}$La$_z$(Ba$_{1-x}$La$_x$)$_2$Cu$_3$O$_y$ and efficiently
removing oxygen using thin-film growth method. In this work, we grow
the Y$_{0.38}$La$_{0.62}$(Ba$_{0.82}$La$_{0.18}$)$_2$Cu$_3$O$_y$
thin films using a pulsed laser deposition (PLD) process and shift
the materials from $p$-type superconducting to $n$-type metallic
states by postannealing in low \emph{P}$_{O2}$. The resistivity of
$n$-type samples exhibits a Fermi-liquid behavior and an anomaly at
a certain \emph{T} below which the resistivity begins to decrease
rapidly. This behavior is similar to that of the underdoped
non-superconducting NCCO, suggesting the potential of $n$-type
superconductivity in YBCO system if electrons are further doped.

\section{Experimental}

The starting materials for the preparation of the ceramic target
were pure cation oxides powders of Y$_2$O$_3$ (99.999 $\%$),
La$_2$O$_3$ (99.999 $\%$), BaCO$_3$ (99.997 $\%$) and CuO (99.9999
$\%$). These were weighed and mixed according to the chemical
formula of
Y$_{0.38}$La$_{0.62}$(Ba$_{0.82}$La$_{0.18}$)$_2$Cu$_3$O$_y$. The
mixture was then fired at 850, 900, 900 $^\circ$C for 10 h
respectively, and at 980 $^\circ$C for 20 h in air with regrinding
between firings. On the final firing, the powder was pressed into
disk-shaped pellet for the PLD process. Thin films with thickness of
$\sim$260 nm were grown on (001) LaAlO$_3$ (LAO) substrates by a PLD
system using the as-prepared target. The deposition \emph{T} and
\emph{P}$_{O2}$ for all samples were 760 $^\circ$C and 200 mTorr,
respectively. Since we cannot measure accurately the oxygen content,
we label the films annealed at different conditions by carriers per
Cu atom which is determined by the carrier densities (from Hall
measurements) and volume of primitive cell of YBCO. Three $p$-type
samples with carrier densities (at 300 K) of \emph{p}=0.019, 0.034,
0.055 holes/Cu were obtained by \emph{in situ} postannealing in the
PLD chamber at 560 $^\circ$C for 20 min in \emph{P}$_{O2}$=0.02, 0.2
and 3 Torr, respectively. $P$-type sample with higher carrier doping
of \emph{p}=0.068 holes/Cu was obtained by re-annealing in a tube
furnace at 600 $^\circ$C for 30 min in air. $N$-type samples with
carrier densities of \emph{n}=0.02, 0.029, 0.034, 0.038 electrons/Cu
were obtained by \emph{in situ} postannealing at 640 $^\circ$C in
vaccum (\emph{P}$_{O2}$$<$10$^{-5}$ Torr) for 10, 30, 50 and 80 min,
respectively, and then cooling down to room temperature at 30
$^\circ$C/min. In order to obtain higher electron doping of
\emph{n}=0.087 and 0.166, the samples annealed for 80 min were
re-annealed in vaccum at 380 $^\circ$C with a ramp rate of 30
$^\circ$C/min for 0 min (with immediate cool down at \emph{T}=380
$^\circ$C) and 10 min, respectively. Note that this re-annealing
process is critical for reduction of oxgen as confirmed  by
expansion of the c-axis (fig. 1 and fig. 5). The composition of the
grown films was analyzed by the Rutherford Backscattering
Spectrometry (RBS) and was found to approximately be the same as
that of the target. The crystallographic structure of the thin films
was measured by X-ray diffraction (XRD). The transport property
measurements were made using a Quantum Design PPMS at temperatures
ranging from 2 to 400 K. The resistivity was measured by four-probe
method and Hall effects by Van der Pauw geometry with the magnetic
field swept from -5 to 5 T.

\begin{figure}
\includegraphics[width=3.2in]{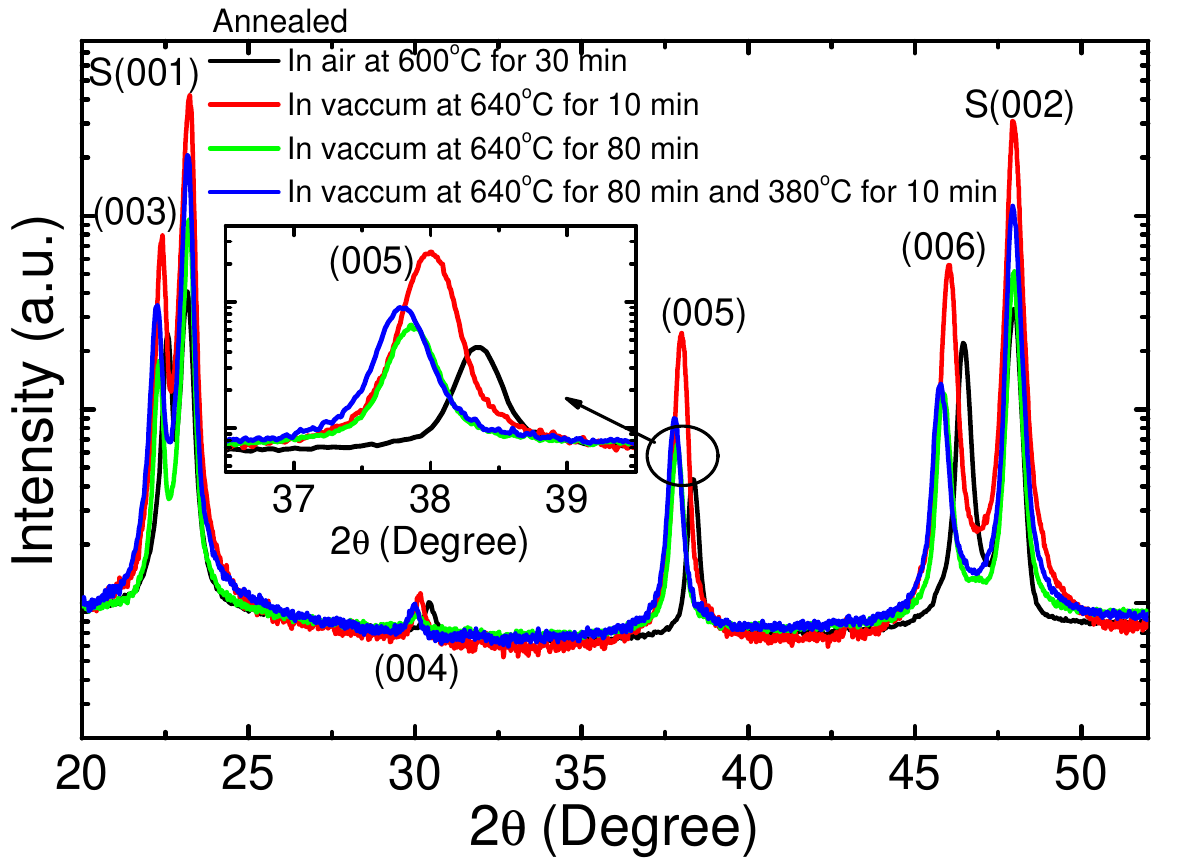}
\caption{\label{fig1} X-ray diffraction patterns of
Y$_{0.38}$La$_{0.62}$(Ba$_{0.82}$La$_{0.18}$)$_2$Cu$_3$O$_y$ thin
films. Inset: (005) peaks.}
\end{figure}

\section{Results and discussion}

Figure 1 shows the $\theta$-2$\theta$ XRD patterns of four
Y$_{0.38}$La$_{0.62}$(Ba$_{0.82}$La$_{0.18}$)$_2$Cu$_3$O$_y$ samples
annealed in air and vacuum. Only (00$\l$) peaks of thin films and
LAO substrates (indicated with S(00$\l$)) were clearly observed,
where $\l$ is an integer, confirming the $c$-axis oriented epitaxial
growth. The $c$-axis lattice constant, \emph{d}, of the sample
annealed in air was determined to be 11.728 $\AA$ which is less than
that of the as-grown (in air) crystal
Y$_{0.38}$La$_{0.62}$(Ba$_{0.87}$La$_{0.13}$)$_2$Cu$_3$O$_y$ (11.763
$\AA$) [8]. This indicates that our samples have larger La
composition substituted for Ba and thus, possess smaller \emph{d},
which is in fair agreement with the established empirical relation
[12]. (00$\l$) peaks of the samples annealed in vacuum shift to
smaller angles, compared with that of sample annealed in air,
suggesting an expansion of the $c$-axis. It is known that \emph{d}
of YBa$_2$Cu$_3$O$_y$ increases as $y$ decreases [13, 14]. This
supports the fact that oxygen was removed as
Y$_{0.38}$La$_{0.62}$(Ba$_{0.82}$La$_{0.18}$)$_2$Cu$_3$O$_y$ films
were annealed in low \emph{P}$_{O2}$ and vaccum.

Figure 2(a)-2(d) show the evolution of the \emph{T} dependence of
$\rho$$_{ab}$ upon changing carrier doping levels at 300 K. The $p$-
and $n$-type charge carriers were confirmed by Hall effects
measurements in fig. 4. Samples with high hole-doping level show
superconductivity, for example, sample with \emph{p}=0.068 exhibits
zero-resistance \emph{T}$_c$ of $\sim$8 K. As \emph{p} decreases,
and thus the oxygen content in the sample is reduced, \emph{T}$_c$
decreases and $\rho$$_{ab}$ increases, the samples show evolution
from superconductors to insulators (fig 2(b)). To further reduce
oxygen content and obtain electron doping, thin films were annealed
in vacuum. For low electron doping level, the sample exhibits
insulating behavior (fig. 2(b)). As doping increases, $\rho$$_{ab}$
goes down and the samples show metallic behavior at high \emph{T}
(fig. 2(c), (d)), although $\rho$$_{ab}$ shows an upturn at low
\emph{T}. For $n$-type samples with \emph{n}$<$0.087, the magnitude
of $\rho$$_{ab}$ within metallic region is above the Mott limit
[15]. Moreover, the metallic behavior is already established at
doping \emph{n}=0.029 (5.0$\times$10$^{20}$ cm$^{-3}$) near Mott
insulating state [9] and electron-electron scatterings are dominant
in transport behavior which will be discussed in the following text.
This suggests a bad metal behavior at low doping [17-19]. However,
the sample with \emph{n}=0.166 is obviously metallic since
$\rho$$_{ab}$ is below the Mott limit from $\sim$300 K down to the
lowest \emph{T} measured. The $\rho$$_{ab}$ of the sample with
highest doping ($\sim$0.88 m$\Omega$$\cdot$cm at 300 K), to the best
of our knowledge, is the lowest resistivity in $n$-type YBCO system
[8-9,11].

\begin{figure}
\includegraphics[width=3.4in]{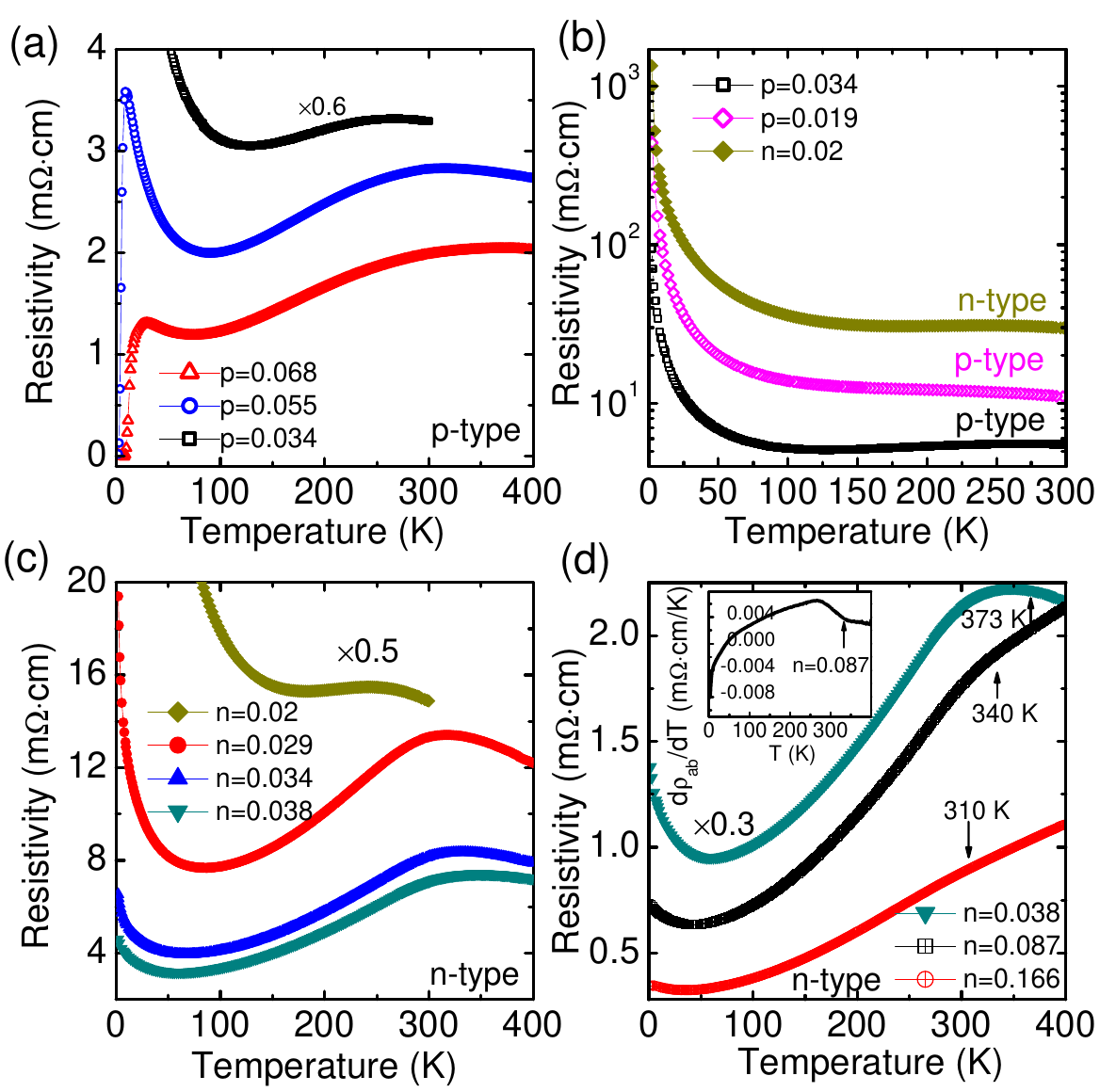}
\caption{\label{fig3} The in-plane resistivity $\rho$$_{ab}$ as a
function of temperature for $p$-type and $n$-type thin films. The
samples were labeled by the hole (\emph{p}) and electron (\emph{n})
doping at 300 K (per Cu atom) which is obtained by Hall measurement.
Inset of (d) is the temperature derivative of $\rho$$_{ab}$ for
sample with \emph{n}=0.087.}
\end{figure}

In the moderate-temperature range, $\rho$$_{ab}$ of the $n$-type
samples is approximately proportional to \emph{T}$^2$, as is
demonstrated in fig. 3, indicating a conventional Fermi-liquid (FL)
behavior due to electron-electron scattering [2]. By fitting the
data to the equation
$\rho$$_{ab}$(T)=$\rho$$_{0}$+\emph{A}$_2$\emph{T}$^2$, we can
obtain the quadratic scattering rate \emph{A}$_2$. \emph{A}$_2$
decreases rapidly from $\sim$8.6$\times$10$^{-5}$ to
$\sim$7.5$\times$10$^{-6}$ m$\Omega$$\cdot$cm$\cdot$K$^{-2}$ with
increasing electron doping, as is shown in fig. 5. This behavior of
\emph{A}$_2$ is similar to those in NCCO as electron doping
increases [20]. Interestingly, the magnitude of \emph{A}$_2$ in our
films are of the order of 10$^{-6}$ $-$ 10$^{-5}$
m$\Omega$$\cdot$cm$\cdot$K$^{-2}$ which is comparable to those of
NCCO [2, 20]. The similarity in magnitude and evolution of
\emph{A}$_2$ between
Y$_{0.38}$La$_{0.62}$(Ba$_{0.82}$La$_{0.18}$)$_2$Cu$_3$O$_y$ and
NCCO hint at the possibility that the electron-electron scattering
in $n$-type cuprates is governed by essentially the same physics,
regardless of the different crystallographic structures. As is
marked by arrows, the FL regimes shift to lower \emph{T} as the
electron doping increases. It has been demonstrated that the ground
state of heavily overdoped $n$-type cuprates is dominated by FL
behavior [21]. The shift of FL regimes probably suggests the same
ground state in $n$-type YBCO system if electrons are overdoped. In
contrast, the quadratic dependence with \emph{T} of $\rho$$_{ab}$ is
not observed in $p$-type samples.

\begin{figure}
\includegraphics[width=3.4in]{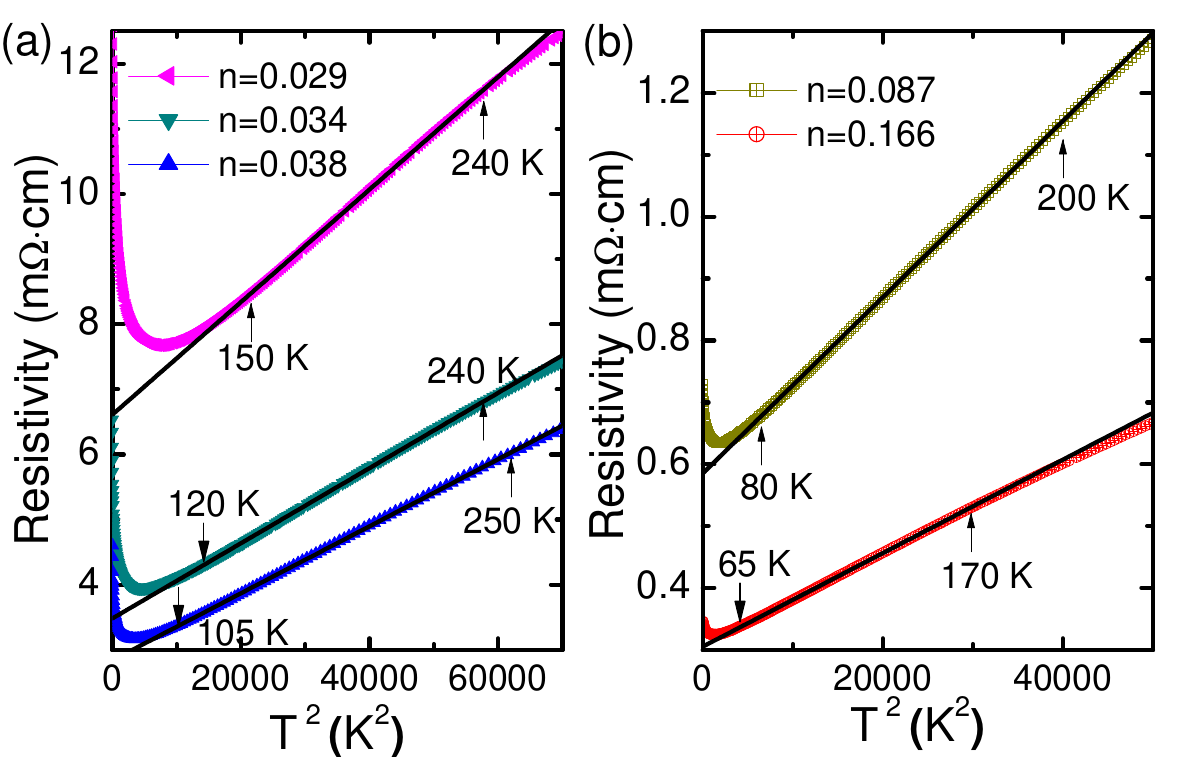}
\caption{\label{fig3} $\rho$$_{ab}$ of $n$-type samples as a
function of \emph{T}$^2$ for the same data shown in fig. 2. Solid
line is the fitting to the data. Arrows indicate the \emph{T} where
$\rho$$_{ab}$ deviates from Fermi-liquid behavior.}
\end{figure}

At \emph{T} above $\sim250$ K, weakening of the quadratic dependence
of $\rho$$_{ab}$ is observed for electron-doped samples [2].
$\rho$$_{ab}$ even saturate and become non-metallic at higher
\emph{T} for \emph{n}$<$0.087. For \emph{n}=0.087 and 0.166,
$\rho$$_{ab}$ tends to saturate at higher \emph{T} although it is
metallic up to 400 K. The saturation of $\rho$$_{ab}$ can also be
observed in $p$-type samples. On close examination, for samples with
$n$$\ge$0.038, $\rho$$_{ab}$ begins to decrease rapidly at around
the \emph{T} marked by arrow (\emph{T}$_{\rho}$) with decreasing
\emph{T}. \emph{T}$_{\rho}$ can be obtained by the \emph{T}
derivative of $\rho$$_{ab}$ and is around the \emph{T} where
d$\rho$$_{ab}$/d\emph{T} starts to increase [22], as is shown in
inset of fig. 2 (d). The steep decrease of $\rho$$_{ab}$ is also
observed in underdoped Nd$_{2-x}$Ce$_x$CuO$_4$ with $x$=0.025-0.075
and is related to the formation of pseudogap confirmed by optical
spectra [22,23]. In Nd$_{2-x}$Ce$_x$CuO$_4$, \emph{T}$_{\rho}$
decreases as $x$ increases and disappears at optimal doping of
$x$=0.15. In our samples, \emph{T}$_{\rho}$ also decreases with
increasing electron doping. Importantly, this anomaly in
$\rho$$_{ab}$ become less noticeable at higher electron doping which
is similar to NCCO. Whether the pseudogap begins to evolve below
\emph{T}$_{\rho}$ requires further investigation by optical
spectroscopy. Nevertheless, the anomalies in $\rho$$_{ab}$ of
$n$-type films suggest the possibility of its being a precursor of
superconductivity.

\begin{figure}
\includegraphics[width=3.4in]{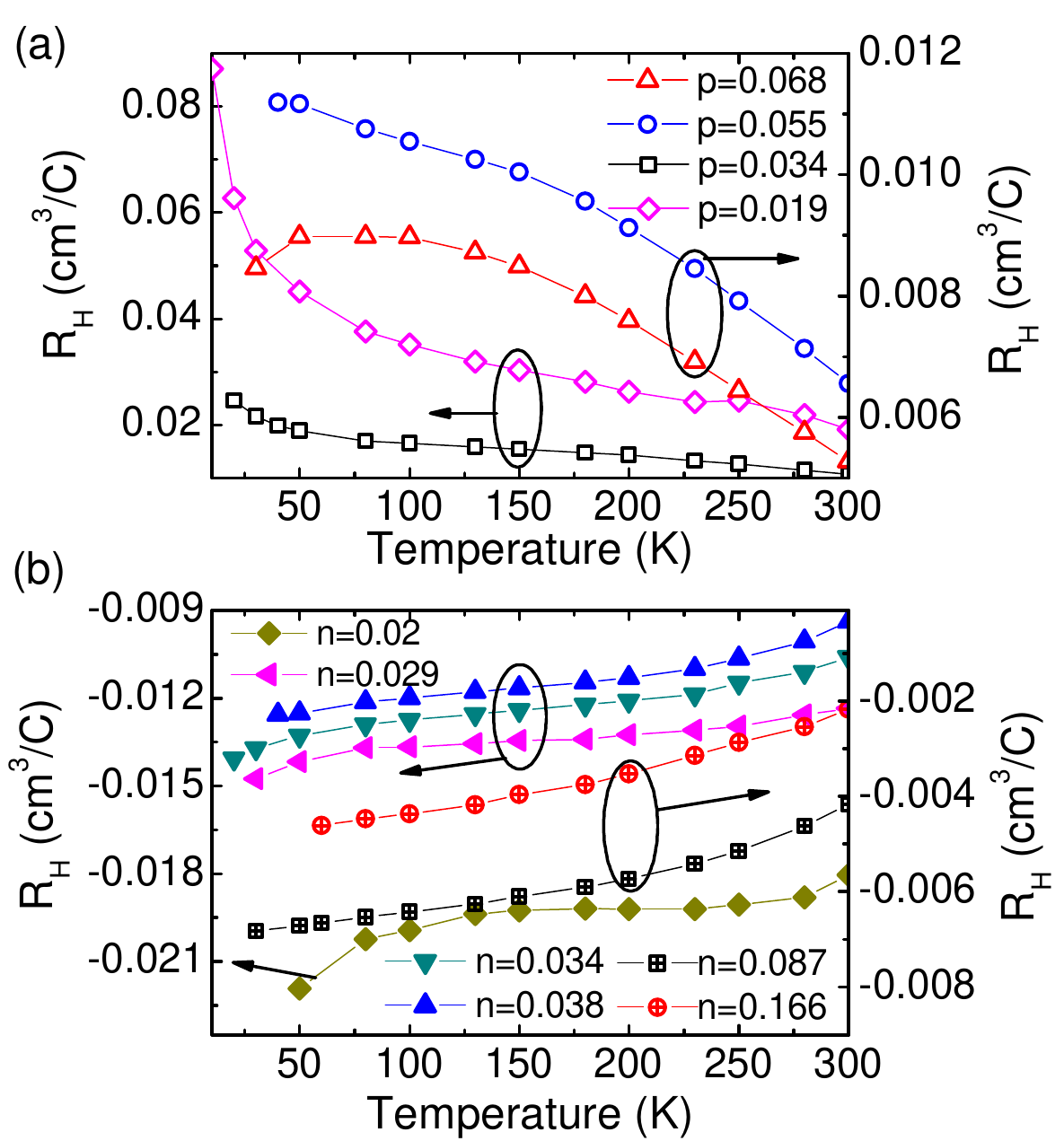}
\caption{\label{fig4} The Hall coefficients \emph{R}$_H$ of $p$-type
(a) and $n$-type (b) thin films as function of temperature.}
\end{figure}

Figure 4 shows the Hall coefficients (\emph{R}$_H$). The samples
annealed in vaccum exhibit negative \emph{R}$_H$ all the way below
300 K, indicating electron doping. The electron density of optimally
reduced sample is $\sim$2.87$\times$10$^{21}$ cm$^{-3}$
(\emph{R}$_H$$\sim$ -0.00217 cm$^3$/C) at 300 K which is one order
of magnitude higher than those in
Y$_{0.38}$La$_{0.62}$(Ba$_{0.87}$La$_{0.13}$)$_2$Cu$_3$O$_y$ single
crystal ($\sim$2.2$\times$10$^{20}$ cm$^{-3}$, \emph{R}$_H$$\sim$
-0.028 cm$^3$/C) [8] and pure YBCO film ($\sim$2.5$\times$10$^{20}$
cm$^{-3}$) [11]. The magnitude of \emph{R}$_H$ in insulating film
with \emph{n}=0.02 (\emph{R}$_H$$\sim$ -0.018 cm$^3$/C at 300 K)
increase more sharply at low \emph{T}, which is similar to that of
the single crystal with near electron doping (\emph{R}$_H$$\sim$
-0.028 cm$^3$/C at 300 K) [8]. Futhermore, $\rho$$_{ab}$ at 300 K of
the insulating thin film exhibits similar value ($\sim$29
m$\Omega$$\cdot$cm) to that in single crystal ($\sim$30
m$\Omega$$\cdot$cm) [8]. These suggest that the crystal quality of
our thin films is comparable to that of single crystal. As is shown
in fig. 5, \emph{d} continuously increases by hole depletion and
electron doping across the zero-doping state, although it is
moderate at higher electron-doping region, which indicates that
electrons were continuously doped as oxygens were removed.
Therefore, high carrier density and low $\rho$$_{ab}$ in $n$-type
thin films are caused by electron doping.

\begin{figure}
\includegraphics[width=3.4in]{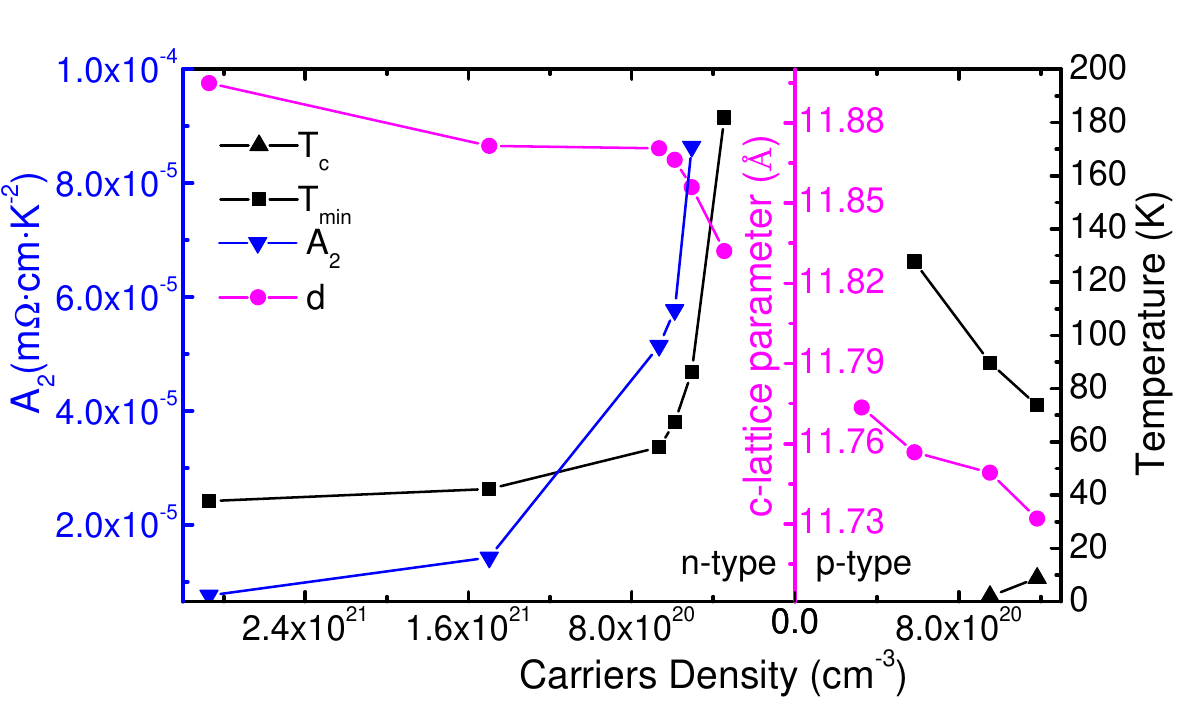}
\caption{\label{fig5} Superconductivity transition temperature
\emph{T}$_c$, temperature with minimum resistivity \emph{T}$_{min}$,
c-lattice parameter \emph{d} and quadratic scattering rate
\emph{A}$_2$ as a function of carriers density at 300 K.
\emph{T}$_{min}$ is obtained by calculating the derivatives
\emph{d}$\rho$$_{ab}$/\emph{dT} and is the temperature when
\emph{d}$\rho$$_{ab}$/\emph{dT}=0. \emph{A}$_2$ is obtained from
fitting $\rho$$_{ab}$(T)=$\rho$$_{0}$+\emph{A}$_2$\emph{T}$^2$ to
data within $\rho$$_{ab}$$\propto$\emph{T}$^2$ region shown in
fig.3.}
\end{figure}

It is found that in all the samples with metallic state, even in
superconducting $p$-type samples, $\rho$$_{ab}$ exhibits
insulating-like behavior below \emph{T} with minimum $\rho$$_{ab}$
(\emph{T}$_{min}$). This behavior have been investigated in
underdoped cuprates but its origin is still unclear [24-26]. We also
show \emph{T}$_c$ and \emph{T}$_{min}$ as a function of carrier
density at 300 K (\emph{n}$_{300K}$) in fig. 5. \emph{T}$_{min}$ was
found to share the same evolution as a function of doping in both
$n$- and $p$-type thin films, mainly decreasing with increasing
\emph{n}$_{300K}$. However, \emph{T}$_{min}$ decreases much more
rapidly in $n$-type samples, from 181.8 K at
\emph{n}$_{300K}$$\approx$3.46$\times$10$^{20}$ cm$^{-3}$ to 57.9 K
at \emph{n}$_{300K}$$\approx$6.65$\times$10$^{20}$ cm$^{-3}$, than
that in $p$-type samples, from 127.7 K at
\emph{n}$_{300K}$$\approx$5.8$\times$10$^{20}$ cm$^{-3}$ to 73.7 K
at \emph{n}$_{300K}$$\approx$1.18$\times$10$^{21}$ cm$^{-3}$. At
higher electron doping, \emph{T}$_{min}$ exhibits a slight drop and
tends to saturate. Superconductivity emerges at
\emph{n}$_{300K}$$\approx$9.5$\times$10$^{20}$ cm$^{-3}$
(\emph{T}$_{c}$=2 K) when \emph{T}$_{min}$=89.7 K in $p$-type
samples and \emph{T}$_{c}$ increases with decreasing
\emph{T}$_{min}$. However, superconductivity was not observed in
$n$-type samples even if \emph{T}$_{min}$ was much lower than 89.7 K
and the doping level of electrons was higher than that of holes.
Interestingly, it was found that the amplitude of \emph{A}$_2$
exhibits the same evolution as \emph{T}$_{min}$ with electron
doping, suggesting an intimate relationship between
electron-electron scattering and the metal-insulator transition in
$n$-type YBCO system.

\section{Conclusion}

In conclusion, metallic $n$-type La-doped YBCO thin films were
successfully obtained by PLD technique and postannealing in vaccum.
Above around \emph{T} where metallic-insulating transition, the
$n$-type samples exhibited \emph{T}$^2$-dependent resistivity. At
relatively higher \emph{T}, $\rho$$_{ab}$ showed an anomaly probably
due to the formation of pseudogap which has been observed in
underdoped NCCO. $P$-type samples were also investigated and showed
different evolutions with doping in $\rho$$_{ab}$(T), \emph{T}$_c$
and \emph{T}$_{min}$. The present work could be a significant step
toward ambipolar superconductivity in YBCO system.

\begin{acknowledgments}
We thank the National Research Foundation (NRF) Singapore under the
Competitive Research Program (CRP) ``Tailoring Oxide Electronics by
Atomic Control'' NRF2008NRF-CRP002-024, National University of
Singapore (NUS) cross-faculty grant and FRC (ARF Grant No.
R-144-000-278-112) for financial support.
\end{acknowledgments}


\end{document}